\documentstyle[12pt,epsfig]{article}

\newcommand{\be}{\begin{equation}}
\newcommand{\ee}{\end{equation}}

 1
\font\elevenrm=cmr10 scaled\magstep 1
 1

\def\reff{\hang\noindent}

\begin{document}
\vspace*{1.8cm}
  \begin{center} {\bf STUDY OF PHOTONUCLEAR INTERACTION OF MUONS IN ROCK \\
WITH THE MACRO EXPERIMENT}
\end{center}
\vspace{1cm}
  \begin{center}
G. BATTISTONI$^a$ for the MACRO\footnote{The MACRO Collaboration: 
M. Ambrosio, 
R. Antolini, C. Aramo, G. Auriemma, A. Baldini, 
G. C. Barbarino, B. C. Barish, G. Battistoni, 
R. Bellotti
% ...
\\
} Collaboration \\
\vspace{1.4cm}
  {\elevenrm a) INFN, Sezione di Milano, via Celoria 16, I-20133 Milano, Italy} \\
\end{center}
\vspace{3cm}
\begin{abstract}
We present first results about the measurement of the
characteristics of charged hadrons production by atmospheric muons in
the rock above MACRO. Selection criteria which allow to discriminate hadron
cascades from e.m. showers generated by muons are described.
A comparison between the measured rate, with that expected from a Monte Carlo 
simulation which treats the process as dominated by photo-nuclear 
interaction is presented.
These data can be used to validate such models aiming to the 
evaluation of hadron background from cosmic muons
in different experimental environments.
\end{abstract}
\vspace{2.0cm}

\section{ Introduction }

The inelastic muon-nucleus interaction was discovered in by George and 
Evans (1955), who observed ``stars'' of charged hadrons produced by 
high energy muons. Since then the process is generally referred to as nuclear
interaction of muons.
At accelerators, this process is mainly studied in the range of large squared 
four-momentum transfer ($Q^{2}\ge 1 GeV^{2}$) with the principal aim of 
measuring nucleon structure functions (deep inelastic scattering experiments).
However, the bulk of interactions are characterized 
by low $Q^{2}$ ($Q^{2}\le 0.1 GeV^{2}$): hence they can be described with
the exchange of a quasi-real photon between the muon and the nucleon and they 
are often referred to as photo-nuclear interaction of muons.
Recently, it has been stressed that nuclear interactions of muons are an
important source of background for many underground experiments
(Khalchukov et al, 1995).
Low energy protons coming either from the primary interaction of muons or
from reinteraction of produced hadrons can be a relevant background
in the detection of solar $\nu_e$'s by radio-chemical means
(Cribier et al., 1997), while neutral hadrons may play a significant role 
in oscillation experiments at reactor or accelerators 
(see Kleinfeller et al., 1996).
Furthermore, the question of hadron background generated by muons has been 
raised in the search of proton decay (Khalchukov et al, 1983)  and
in the study of atmospheric neutrinos, observed as contained 
events in Cherenkov detectors (Ryazhskaya, 1994; Becker-Szendy et al., 1992), 
or as upward going muons 
produced by CC interactions in the rock (MACRO Coll., 1997a).

A comparison between experimental measurements about the production of hadrons
by muons and data obtained by Monte Carlo (MC) simulations is mandatory to test the 
reliability of theoretical models.
In fact, uncertainties do exist on the cross section calculation for
the process, mainly due to the extrapolation of photo-nuclear 
cross section at high energies, and in the 
simulation 
of the hadronic final state, because the smallness of the $Q^{2}$ does not 
allow the use of perturbative QCD, and models are required.
These uncertainties also affect muon survival probability calculations in 
underground-underwater experiments (Kokoulin and Petrukhin, 1996).

The aim of the present work is to compare real data collected by the large area
underground experiment MACRO (MACRO Coll., 1993), 
operating in the underground Gran Sasso laboratory,
 with a MC simulation based on the cross section calculated by Bezrukov and 
Bugaev (1981) and on DPM model (see for instance Capella et al, 1987) 
for the sampling of the 
final state. We shall look for comparison also to a complete different
modelization available in the high energy physics library.

In next sections we briefly review the detector features and
the model employed to calculate the 
cross section used in our simulation. Then we discuss
the analysis and the preliminary results.

\section{The Detector}

A description of MACRO and in particular of his tracking system 
is given elsewhere (MACRO Coll., 1993, and references
therein).  We limit to remind here that MACRO operates at an average depth
of 3800 hg/cm$^2$, where the average residual energy of muons is
about 300 GeV. The
horizontal area of MACRO is about 1000 m$^2$ and it
can reconstruct charged tracks, by means of a streamer tube system, 
in two projective views, with 
a space point accuracy of $\sim$1 cm and an angular accuracy
better than 1 degree.
The total thickness of the detector is such that the probability of detecting and
containing a photonuclear interaction is small, and it is more convenient
to study the interactions in the surrounding rock. The upper part of MACRO, in particular,
is in practice a thin detector, imposing a low threshold on secondary particles.
The lower part, thanks to the rock absorbers, allows to stop particles and measure their
range up to few hundreds of MeV.

\begin{figure}[!t]
\begin{center}
\begin{tabular}{c}
\includegraphics[width=7.1cm,height=7cm]{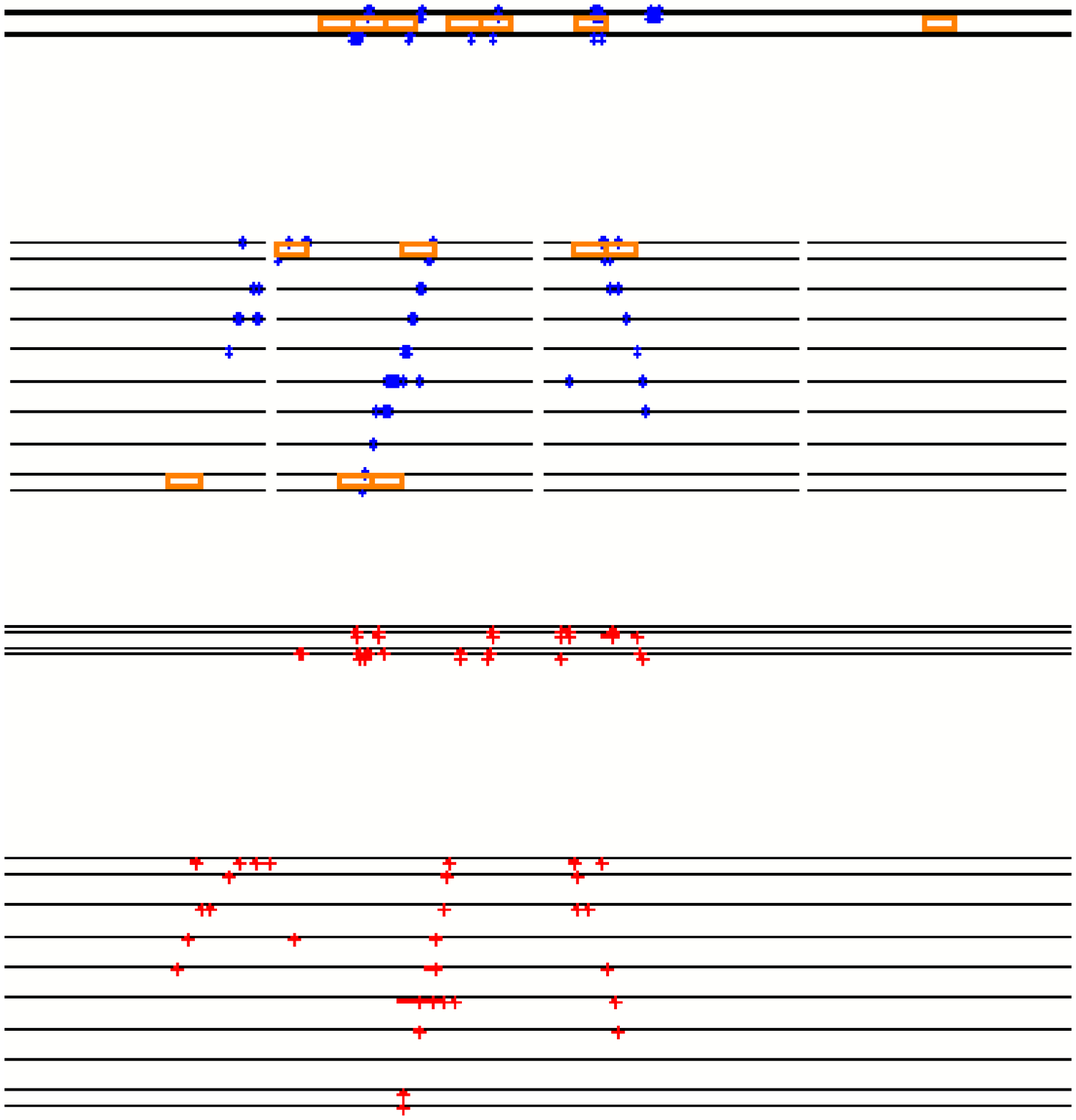} \\
\includegraphics[width=7cm]{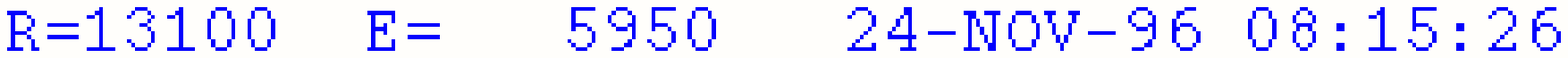}
\end{tabular}
\caption{A typical candidate event as detected in MACRO.
The two different projective views are shown one above the other.
\label{bellevento}}
\end{center}
\end{figure}

We consider events in which the 
muon interacts in the rock above the apparatus and both the muon and at least
one charged hadron are observed in the horizontal planes of
the tracking detector. An example is shown in Fig.~\ref{bellevento}, 
in which a muon enters the apparatus from above,
along with at least one charged hadron.
We have not the possibility of  an efficient
detection of the  neutrons produced in the interaction: yet,
hadronic cascades generated by inelastic collisions of multi GeV neutrons 
inside the rock absorber layers of our apparatus are observed.

With the aim of recognizing the muon, the standard MACRO muon tracking is used,
selecting tracks entering from the uppermost plane, and with at least 7 hit 
horizontal planes and $cos(Zenith) \ge 0.4$. 
The additional tracks have been recognized by means of a specialized
algorithm, looking for shorter tracks pointing towards the main track
around a common vertex region contained in the rock above the detector.
In practice we select charged hadrons with a minimum kinetic energy around 
150 MeV.

\section {The physical process and its simulation}

\begin{figure}[!tp]
\begin{center}
\includegraphics[width=11cm]{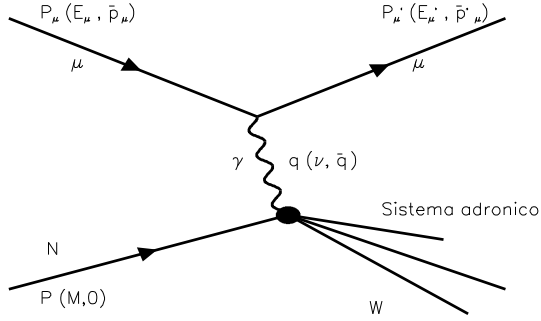}
\caption{Lowest order diagram of the photonuclear process.}
\label{Fey_d}
\end{center}
\end{figure}

The process can be represented by the 
Feynman diagram shown in Fig.~\ref{Fey_d}, in which
$P_{\mu}$ and $P'_{\mu}$  are the four-momenta of the muon
before and after the interaction respectively (their components in laboratory 
reference system are indicated); $q=P_{\mu}-P'_{\mu}$ is the four-momentum 
transferred to the nucleon by the photon; $P$ and $W$ are the four-momenta
of the nucleon before interaction (supposed at rest, we shall neglect Fermi motion
in the nucleus), 
and of the final state hadronic system.
From the calculation point of view, the emission of the
photon can be treated in the 
Williams-Weizsacker approximation, in which the passage of a charged 
lepton in a slab of material produces the same effects of a beam of quasi-real
photons.
As far as interaction of a muon with a nucleon at low $Q^{2}$ is 
concerned, the most used form for the differential cross section is the 
following:\\
\newcommand{\ud}{\mathrm{d}}
\newcommand{\alphpi}{\frac{\alpha}{\pi}\,}

\begin{equation} \label{sig:q2:v}
\frac{\ud^{2} \sigma_{\mu-N}}{\ud v\,\ud Q^{2}}=
  \alphpi (\Gamma_t\,\sigma_t + \Gamma_l\,\sigma_l)=
   \alphpi \Gamma_t\,(\sigma_t + \epsilon\,\sigma_l)=\alphpi
    \Gamma_t\,\sigma_{virt}(Q^2,\nu)
\end{equation}
where $\sigma_t$ and $\sigma_l$ are the cross sections for the interaction of
transverse and longitudinal photons with a nucleon, 
$\epsilon=\frac{\Gamma_l}{\Gamma_t}$ is the polarization factor an depends upon
$E,\nu,Q^2$. In the Williams-Weizsacker approximation, $\Gamma_t$ is related to
the energy spectrum of equivalent photon beam.
Both
$\sigma_l$ and $\sigma_t$, and hence $\sigma_{virt}$ in Eq.~\ref{sig:q2:v} 
are closely related to $\sigma_{\gamma-N}(\nu)$, the cross section for the 
interaction between a real photon with energy $\nu$ and a nucleon. 
In the low $Q^2$ region, $\sigma_{virt}$ can be 
expressed in the form $\sigma_{virt}=\sigma_{\gamma-N}(\nu)\,F(Q^2)$, 
where $F(Q^2)$ is the nucleon structure form factor.
When the interaction with a \emph{nucleus} with mass $A$ is considered, 
the ``shadowing'' effect has to be taken into account: this effect is expressed by
the fact that $\sigma_{\mu-A}$ is 
somewhat less than the mere $A\,\sigma_{\mu-N}$.
In the framework of the ``Vector Meson Dominance'' (VMD)
the photon radiated by the muon
interacts with nucleons in a nucleus by virtually converting in a vector
mesons (mostly $\rho$). Thus, by using the optical model for hadron-hadron
interaction, shadowing is explained as due to destructive interference 
of scattering amplitudes. 
By using the generalized
version of this model (GVD), Bezrukov and Bugaev (1981) calculated
the differential cross section for photon-Nucleus and hence for 
muon-Nucleus interactions. 
In the FLUKA Monte carlo code (Fass\'o et al, 1997)
, this process has been implemented, following
the quoted Bezrukov-Bugaev model.
The algorithm in FLUKA for the photonuclear interaction
is realized according to the following steps.
Photon energy $E_\gamma = E_\mu\,v$ is sampled according to 
$\frac{\ud \sigma}{\ud v}$ given by Bezrukov and Bugaev (1981).
Then, following VMD, the photon is coupled to a vector meson
($\rho$, $\omega$, and $\phi$) with the known branching ratios.
Whenever photon energy is too small, it is treated like a pion.
An on shell mass is given to $V$ according to its observed width.
The interaction is treated in the $\gamma$-nucleon centre of momenta system, 
and the final state is sampled according to the Dual Parton Model at
high energies, or to a cascade pre-equilibrium model at low energy.  
Deep inelastic scattering of muons on nucleons is not yet
included in FLUKA. As already mentioned, this should not be a great problem
because the region of low $Q^2$ gives the main contribution to the cross 
section. 
 
We have chosen this code as a main reference to generate the underground muon events, 
taking into account the following steps.
The direction of incident muons is sampled according to the local angular 
distribution measured by MACRO and properly unfolded to take into account
the anisotropic acceptance of the apparatus.
As far as muon residual energy at a depth $h$ (in km~w.e.) is 
concerned, we chose 
to sample it according to the approximate distribution (Gaisser, 1990)
following from a 
simple power muon spectrum at surface 
$\Big(\frac{\ud N (E_\mu, 0)}{\ud (E_\mu)} = K\,E_\mu ^{-\gamma}\Big)$ 
and a mean energy loss given 
by: $-\frac{\ud E_\mu}{\ud x} = a + b\,E = a\,(1+\epsilon\,E)$:\\
\begin{equation}
\label{energy}
\frac{\ud N (E_\mu, h)}{\ud (E_\mu)} = K\, e^{-b\,h (\gamma - 1)}
\,(E_\mu + \epsilon \,(1-e^{-b\,h}))^{-\gamma}
\end{equation}
where $\gamma = 3.5$, $b = 4\cdot10^{-6} cm^2/g = 0.4 (km\, w.e.)^{-1}$ and
$\epsilon = a/b \approx 540$ GeV.

Given the direction, the slant depth of rock $h$ crossed by the muon can 
be obtained from the map of the mountain overburden.
In the simulation, muons are allowed to interact in a $13$~m thick layer of
rock positioned all around the experimental hall. This thickness corresponds to about 
\mbox{$35$ interaction lengths} for hadrons, and this is enough to fully
contain hadronic showers: interactions outside this region are practically
invisible because of the ranging out of all the particles possibly
produced.
The actual compound mixture measured at the level of the underground
laboratories has been considered in the simulation.
%
If a photo-nuclear interaction occurs in the region of rock described above,
the muon and secondary particles are transported through the rock, along with
e.m. and hadronic showers possibly produced. 
If both the muon and at least one additional particle reach the tunnel, the 
event is stored.
Furthermore, if the direction of muon happens to cross at least four planes
of MACRO tracking detector, a full simulation of the apparatus response is 
performed by a GEANT 3.21 based package (Brun et al., 1992), 
where, as far as hadron interactions are concerned, 
GEANT-FLUKA interface has been used.
GEANT cuts for $e^\pm, \gamma$ were set to $E^\gamma_{cut}=100 KeV$, while 
1~MeV was used for charged hadrons and 10 MeV for neutrons. 
The simulated data are treated with the
the same analysis instruments of real 
data.

\begin{figure}[htb]
\begin{center}
\includegraphics[width=7cm,height=7cm]{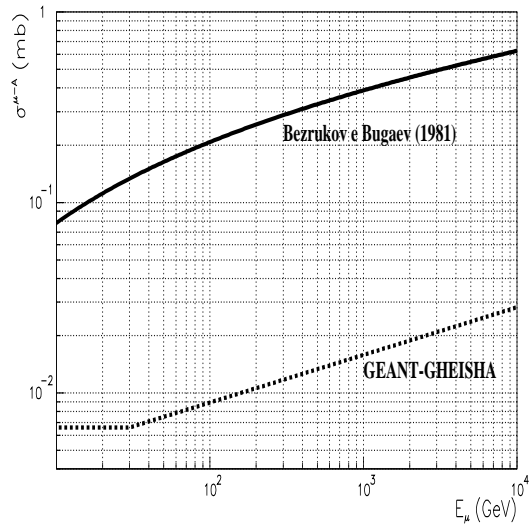} 
\caption{The total cross section for the photonuclear interaction of muon
in rock as a function of muon energy in the two considered models.)
\label{cross}}
\end{center}
\end{figure}

In order to make a comparison with a different model, we have repeated
the event generation in the rock with the photonuclear code of 
GHEISHA code (Fesefeldt, 1985) inside GEANT.
No other simulation of this particular process is available in the present
GEANT environment. One of the most important differences is in the total cross
section, as shown in Fig.~\ref{cross}.
For a more detailed discussion of the differences between FLUKA and GEANT-GHEISHA
for this process, see Battistoni et al., (1997).

\section { Data Analysis}

\begin{figure}[thb]
\begin{center}
\includegraphics[width=15.5cm]{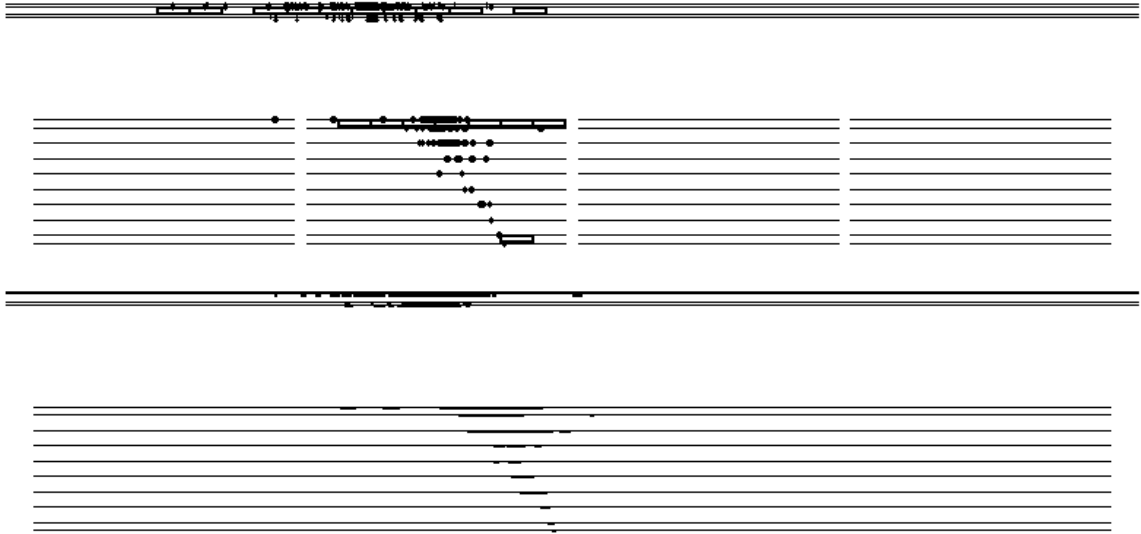} 
\caption{Example of background event: e.m. interaction of muon
in the rock. \label{fi:back1}}
\end{center}
\end{figure}

\begin{figure}[thb]
\begin{center}
\includegraphics[width=15.5cm]{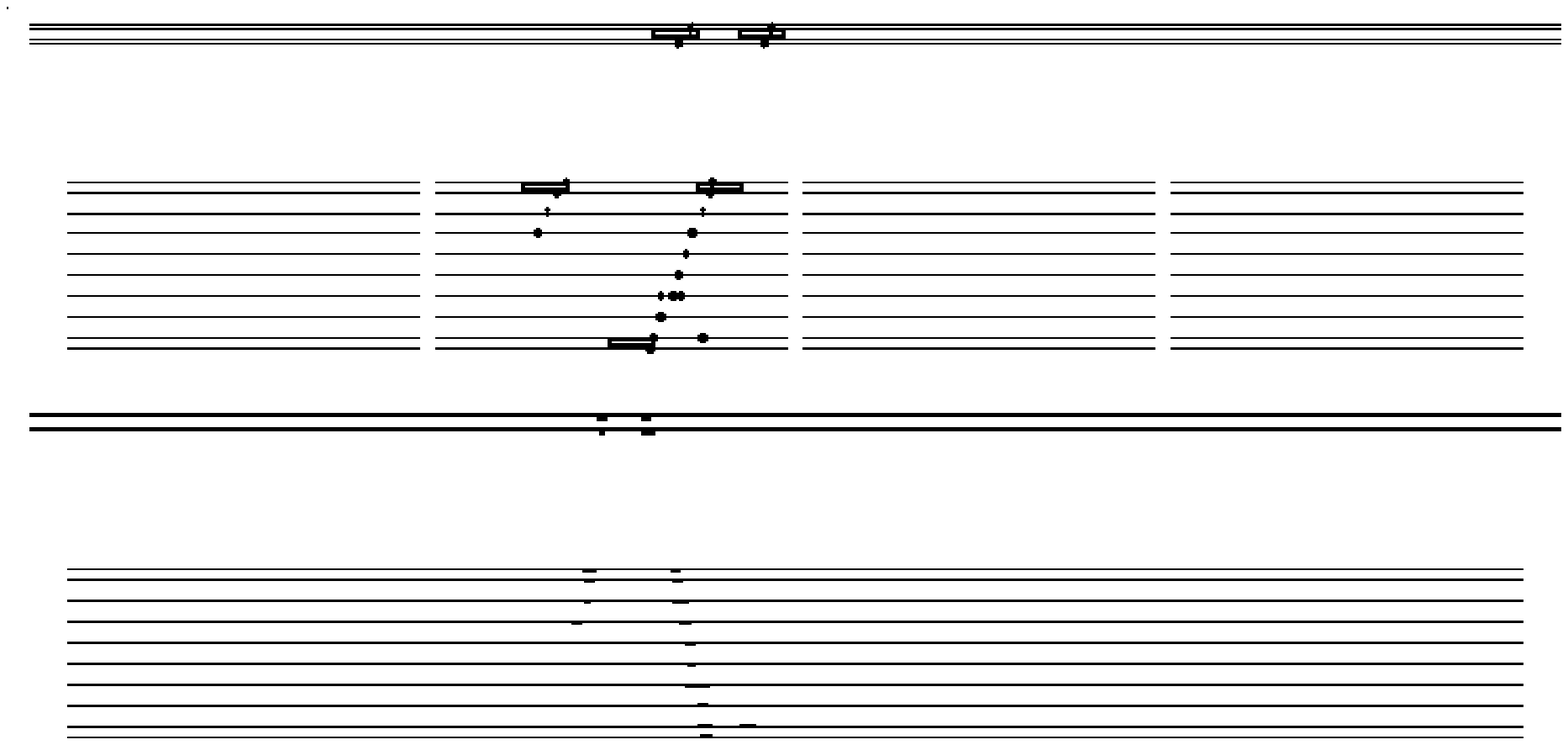}
\caption{Example of background event: fake event
from the muon bundle sample.\label{fi:back2}}
\end{center}
\end{figure}

The main difficulty in our analysis is to achieve
the necessary rejection factor against the physical background.
Such a background is largely dominated by two processes: 
1) the e.m. interactions
of muons in the rock (bremsstrahlung and pair production, which
have a much larger cross section than the photonuclear reaction);
2) multiple muon events in which one of the muons stops inside
the detector.
Examples of background events are shown in Fig.~\ref{fi:back1}
and \ref{fi:back2}

This background has been extensively studied with our simulation
tools. In the case of the e.m. interactions, 
the mean angular separation between the muon and the additional tracks 
is less than the corresponding separation observed in photonuclear
interactions. Besides that, e.m. events often show very large clusters 
of fired tubes near the muon track.
These features are used to achieve a rejection factor against e.m. events
at the level of 4.5$\cdot$ 10$^{-6}$, 
maintaining the recognition efficiency for
hadronic events at the level of 55\%.

In order to reject the muon bundle backgrounds, additional cuts
based on the parallelism of tracks have to be considered.
We have studied the events generated with the HEMAS code 
(Forti et al., 1990)
using for the primary cosmic ray spectrum and mass composition
the results from the best fit reproducing the MACRO data themselves
(MACRO Coll., 1997b).
The rejection factor achieved
for muon bundle background at the level of 1.5$\cdot$10$^{-4}$,
at the price of a slight reduction of the selection efficiency,
which is now at 47\%.

\section { Results}

We have analysed a data sample corresponding
to about 11000 hours of full running of the detector.
With the above selection criteria, we have found
1938 candidate events over a total sample of
9544318 muon events.
From our knowledge of the background, we expect
that our candidates are contaminated by 11 events from the
e.m. interactions in the rock, and by 107 events muon bundles
surviving the cuts.
We can express the results in terms of the ratio 
$R_{\mu+h}$ of the selected $\mu$+hadrons events (background subtracted)
to the number of muon events in the same time.
We then compare the experimental results to the MC prediction having
used the same selection criteria.
After the subtraction of the background, we find for $R_{\mu+h}$ 
in real data and in the MC simulations the following results:

\begin{itemize}
\item $R_{\mu+h}(DATA)=(1.91 \pm 0.05_{stat}\pm 0.03_{syst})\cdot 10^{-4}$, 
\item $R_{\mu+h}(MC-FLUKA)=(1.89 \pm 0.16_{stat}\pm 0.02_{syst})\cdot 10^{-4}$. 
\item $R_{\mu+h}(MC-GEANT/GHEISHA)=(1.31 \pm 0.14_{stat}\pm 0.02_{syst})\cdot 10^{-5}$. 
\end{itemize}
The systematic error on the experimental data is due to the uncertainties on
background subtraction, while the systematic error on the simulation is
dominated by the uncertainties on the muon energy spectrum.

From this preliminary measurement, we can conclude that the MACRO experiment 
can perform the measurement of the charged hadron (E$_{kin}>$ 150 MeV)
production in the rock at the desired level of accuracy.
The FLUKA predictions, based on the Bezrukov and Bugaev model of photonuclear
interaction are in very good agreement with data, while the GEANT-GHEISHA
model gives absolute predictions lower by an order of magnitude.

\section { References}

\reff Battistoni G., et al., 1997: Nucl. Instr. and Methods in Phys. Res. {\bf A394} 136.

\reff Becker-Szendy R., et al., (IMB Coll.), 1992: Phys. Rev. {\bf D46}(9) 3720.

\reff Bezrukov, L. B. and Bugaev E. V., 1981: Sov. J. Nucl. Phys, {\bf 33}(5) 6.

\reff Brun R., et al., 1992: CERN GEANT User's guide, DD/EE 84-1 

\reff Capella A., et al., 1987:  Phys. Rev. Lett. {\bf 58} 2015.

\reff Cribier M., et al., (GALLEX Coll.), 1997: Astrop. Phys. {\bf 6} 129.

\reff Fass\'o A', et al., 1997: \emph{Proc. of the $3^{rd}$ Workshop on
Simulating Accelerator Radiation Environment}, SARE-3, KEK-Tsukuba, 
KEK report 97-5 p. 32. 

\reff Fesefeldt H., 1985: Aachen preprint PITHA 85/02 (1985)

\reff Forti C., et al., 1990: Phys. Rev. {\bf D42} 3668.

\reff Gaisser T. K., 1990: {\it Cosmic Rays and Particle Physics}, Cambridge University Press,
Cambridge, England, Chapt. 6.

\reff George E. P., and Evans J, 1955: Proc. Phys. Soc. {\bf 68} 829.

\reff Khalchukov F. F., et al., 1983: Il Nuovo Cimento {\bf C6}(3) 320.

\reff Khalchukov F. F., et al., 1995: Il Nuovo Cimento {\bf C18}(5) 517.

\reff Kleinfeller J., et al., (KARMEN Coll.), 1996:  Nucl. Phys. {\bf B48} (Proc. Suppl.) 207.

\reff Kokoulin R. P., and Petrukhin A. A., 1996: {\it Proceedings of the 1996 Vulcano Workshop}, 379.

\reff MACRO Coll., 1993: Nucl. Instr. and Methods in Phys. Res. {\bf A324} 337.

\reff MACRO Coll., 1997a: preprint INFN AE-97/55, to be published in Astrop. Phys.

\reff MACRO Coll., 1997b: Phys. Rev. {\bf D56} No. 3 1418.

\reff Ryazhskaya O. G., 1994: JETP lett., {\bf 60}(9) 619.
\end{document}